\documentstyle [12pt]{article}
\normalbaselines

\begin{document}

\def\singlespace {\smallskipamount=3.75pt plus1pt minus1pt
                  \medskipamount=7.5pt plus2pt minus2pt
                  \bigskipamount=15pt plus4pt minus4pt
                  \normalbaselineskip=12pt plus0pt minus0pt
                  \normallineskip=1pt \normallineskiplimit=0pt
                  \jot=3.75pt {\def\smallskip
                  {\vskip\smallskipamount}} {\def\medskip
                  {\vskip\medskipamount}} {\def\bigskip
                  {\vskip\bigskipamount}}
                  {\setbox\strutbox=\hbox{\vrule height10.5pt
                  depth4.5pt width 0pt}} \parskip 7.5pt
                  \normalbaselines} \def\middlespace
                  {\smallskipamount=5.625pt plus1.5pt minus1.5pt
                  \medskipamount=11.25pt plus3pt minus3pt
                  \bigskipamount=22.5pt plus6pt minus6pt
                  \normalbaselineskip=22.5pt plus0pt minus0pt
                  \normallineskip=1pt \normallineskiplimit=0pt
                  \jot=5.625pt {\def\smallskip
                  {\vskip\smallskipamount}} {\def\medskip
                  {\vskip\medskipamount}} {\def\bigskip
                  {\vskip\bigskipamount}}
                  {\setbox\strutbox=\hbox{\vrule height15.75pt
                  depth6.75pt width 0pt}} \parskip 11.25pt
                  \normalbaselines} \def\doublespace
                  {\smallskipamount=7.5pt plus2pt minus2pt
                  \medskipamount=15pt plus4pt minus4pt
                  \bigskipamount=30pt plus8pt minus8pt
                  \normalbaselineskip=30pt plus0pt minus0pt
                  \normallineskip=2pt \normallineskiplimit=0pt
                  \jot=7.5pt {\def\smallskip {\vskip\smallskipamount}}
                  {\def\medskip {\vskip\medskipamount}} {\def\bigskip
                  {\vskip\bigskipamount}}
                  {\setbox\strutbox=\hbox{\vrule height21.0pt
                  depth9.0pt width 0pt}} \parskip 15.0pt
                  \normalbaselines}

\begin{flushright}
{\large  UMD-PP-99-01}
\end{flushright}

\begin{center}
{\large \bf IMPLICATIONS OF THE SUPERKAMIOKANDE RESULT ON THE NATURE OF
NEW PHYSICS}\footnote[1]{Based on the talk presented at the
Neutrino-98 Conference, held at Takayama, Japan, June 4-9, 1998.}

\vspace{1.5cm}

Jogesh C. Pati\footnote[2]{Email:  pati@physics.umd.edu}\\ [5mm]
{Department of Physics\\
University of Maryland\\
College Park, MD 20742  USA\\
(June 20, 1998)}\\[5mm]

\vspace{1cm}

\begin{abstract}
\singlespace
{\small
It is remarked that the SuperKamiokande (SK) discovery of $\nu_\mu$ to
$\nu_\tau$ (or $\rm \nu_X$)-oscillation, with a $\rm \delta m^2 \approx
10^{-2} - 10^{-3} eV^2$ and $\rm sin^2 2 \theta > 0.8$, provides a
clear need for the right-handed (RH) neutrinos. This in turn reinforces the
ideas of the left-right symmetric gauge structure $\rm SU(2)_L
\times SU(2)_R$ as well as SU(4)-color, for which the RH neutrinos are a
compelling feature. It is noted that by assuming
(a) that B-L and $\rm I_{3R}$, contained in a string-derived $\rm
G(224) = SU(2)_L \times SU(2)_R \times SU(4)^c$ or SO(10), break near
the GUT-scale, as opposed to an intermediate scale, (b) the see-saw
mechanism, and (c) the SU(4)-color relation between the Dirac mass of
the tau neutrino and $\rm m_{top}$, one obtains a mass for $\rm 
\nu^\tau_L$ which is just about what is observed.  This is assuming
that the SK group is actually seeing $\rm \nu^\mu_L - \nu^\tau_L$ (rather
than $\rm \nu_L^\mu - \nu_X$)-oscillation.  Following a very recent work
by Babu, Wilczek and myself, it is furthermore noted that by
adopting familiar ideas of understanding Cabibbo-like mixing angles in
the quark-sector, one can quite plausibly obtain
a large $\rm \nu_L^\mu-\nu_L^\tau$ oscillation angle, as observed, in
spite of highly non-degenerate masses of the light neutrinos:  e.g. with $\rm
m(\nu_L^\mu)/m(\nu_L^\tau) \approx 1/10 - 1/20$.  Such non-degeneracy is
of course natural to see-saw. In this case, $\rm \nu^e_L - \nu^\mu_L$ 
oscillation can be relevant to the small angle MSW explanation of the solar
neutrino-puzzle. Implications of the mass of $\rm \nu^\tau_L$ suggested by
the SK result, on proton decay are noted.  Comments are made at the end 
on how the SuperKamiokande result supplements the LEP result in selecting 
out the route to higher unification.
}
\end{abstract}
\end{center}
\newpage
\middlespace

{\bf 1.  Introduction:}  The SuperKamiokande (SK) result, convincingly showing
the oscillation of $\nu_\mu$ to $\nu_\tau$ (or $\rm \nu_X$), with a value of
$\rm \delta m^2 \approx 10^{-2}$ to $\rm 10^{-3}$ $\rm eV^2$ and $\rm sin^2
2\theta > 0.8$\cite{1},
appears to be the {\it first clear evidence} for the existence of new physics
beyond the standard model.  The purpose of this note is to make
two points regarding the implications of the SK result, which though
simple, seem to be far-reaching.  The first is the argument as to why
one needs new physics beyond the standard model.  The second is the
remark that the SK result already tells us much about the nature of the
new physics.  In particular, it seems to suggest clearly the existence
of right-handed neutrinos, a new form of matter,
accompanying the observed left-handed ones.  This in turn
reinforces the twin ideas of the left-right symmetric gauge
structure $\rm SU(2)_L \times SU(2)_R$ and of SU(4)-color,
which were proposed some time ago as a
step towards higher unification \cite{2}.  Either one of these
symmetries require the existence of the right-hand neutrinos.
I note that by assuming (a) that B-L and $\rm I_{3R}$, contained in a
string or a GUT-derived $\rm G(224) = SU(2)_L \times SU(2)_R \times
SU(4)^c$, break near the GUT-scale as opposed to an intermediate or a
low-energy scale, (b) the see-saw mechanism \cite{3}, and (c) the
SU(4)-color relation between the Dirac mass of $\nu_\tau$ and $m_{top}$,
one obtains a mass for $\rm \nu^\tau_L$ which is just about what is
observed.  This is presuming that the SK group is actually
observing $\rm \nu^\mu_L - \nu^\tau_L$, (rather than $\rm \nu^\mu_L -
\nu_X$), oscillation and that the neutrino masses are hierarchical
$\rm (m(\nu^e_L) << m(\nu^\mu_L) << m(\nu^\tau_L))$, so that the observed
value of $\rm \delta m^2$ in fact represents the $\rm (mass)^2$ of $\rm
\nu^\tau_L$.  Such a hierarchical pattern, as opposed
to near degeneracy of two or three neutrino flavors, is of course
naturally expected within the see-saw formula.  Following a very recent
work by Babu, Wilczek and myself \cite{4}, I furthermore note that by
combining contributions to the oscillation angle from the neutrino and
the charged lepton-sectors, and by following familiar ideas on the
understanding of Cabibbo-like mixing angles in the quark-sector, one can
quite plausibly obtain a large $\rm \nu^\mu_L - \nu^\tau_L$-oscillation
angle, as observed, in spite of hierarchical masses of the light
neutrinos:  e.g. with $\rm m(\nu^\mu_L)/m(\nu^\tau_L) \approx 1/10 -
1/20$.  In this case, $\rm \nu^e_L-\nu^\mu_L$ oscillation can be
relevant to the small angle MSW explanation of the solar neutrino
puzzle.  The results on $\rm \delta m^2$
and mixing obtained in the context of G(224) can of course be obtained
within \underline{any extension} of G(224), such as SO(10) \cite{5},
together with supersymmetry.  At the end, implications of the neutrino
mass-scale observed at SuperKamiokande on proton decay are noted.
Comments are made on how the SK result supplements that of LEP in
selecting out the route to higher unification.

{\bf 2.  The Need for New Physics:}  First, as we know, the standard
model (SM), based on the gauge symetry $\rm SU(2)_L \times U(1)_Y \times
SU(3)^C$, contains 15 two-component objects in each family -- e.g. for
the electron-family they are: [$\rm {Q = (u_L, d_L), L = (\nu^e_L,
e^-_L), u_R, d_R \; and \; e_R}$] -
and the Higgs doublet $H = (H^+, H^o)$.
Notice that in the standard model, the left-handed neutrino $\rm \nu_L$
is an \underline{odd ball} in that it is the only member in each family
which does not have a right-handed counterpart $\rm \nu_R$.
This feature in fact carries over to its grand unifying extension $\rm
SU(5)$ as well \cite{6}.  In other words, the standard model (as also
SU(5)) provides a clear distinction between left and right, in the
spectrum as well as in the gauge interactions, and thus explicitly
violates parity and charge conjugation.

Can the neutrinos acquire masses in the standard model?
Without a right-handed counterpart, a left-handed neutrino
$\rm \nu_L$ cannot acquire a Dirac mass.
But it may still acquire a Majorana mass (like
$\rm m_L \nu_L^T  C^{-1} \nu_L)$, by utilizing the effects of quantum gravity,
which of course exists even for the SM, and which may induce a
lepton-number violating non-renormalizable operator (written
schematically) in the form\cite{7}
\begin{equation}
\rm \lambda_L \, LLHH/M_{P{\it l}} + hc.
\end{equation}
Here, $\rm M_{P{\it l}}$ denotes the reduced Planck mass = $\rm 2 \times
10^{18}$ GeV and $\rm \lambda_L$ is the effective dimensionless
coupling.  Apriori, we would expect $\rm \lambda_L$ to be of order one,
unless there are symmetries that are respected by quantum gravity, like
local (B-L),
which may suppress $\rm \lambda_L$; in this case, it
would be \underline{less than} one.  In the SM, however, there is no
such symmetry.  Using the VEV of $\rm <H> \approx 250 GeV$, such an operator
would then give:
\begin{equation}
{\rm m(\nu_L) \approx \lambda_L \frac{(250 GeV)^2}{2 \times 10^{18} GeV}
\approx (\lambda_L)(3 \times 10^{-5} eV)}
\end{equation}

Such a mass would lead to values of $\rm \delta m^2$ (for any two
light neutrino-species) $\rm \leq \lambda^2_L (10^{-9} eV^2).$
This is far too small (even for ridiculously large $\rm \lambda_L
\sim 10^2$, say) compared to the observed value of $\rm \delta m^2 \approx
10^{-2} - 10^{-3} eV^2$.\footnote[1]{One might have asked whether the
mass-scale in the denominator of eq. (1) could plausibly be the GUT
scale ($\rm \approx 2 \times 10^{16} GeV$),
instead of the reduced Planck mass.  That
would have given $\rm m(\nu_L) \approx \lambda_L (3 \times 10^{-3} eV)$,
which is closer but still a bit low compared to the SuperKamiokande
value of $\rm (10^{-1} \; to \; 3 \times 10^{-2} eV)$, unless $\rm \lambda_L
\approx$ 30 to 10.  But, more to the point, in the context of the
standard model, supplemented by just gravity, while Planck mass seems to
have every reason to appear in eq. (1), there does not seem to be any
simple reason for the relevance of the GUT scale.  Putting it another way,
if the GUT scale is needed in eq. (1) for numerical agreement, that by itself
calls for new physics beyond the Standard model.  I thank S. Weinberg, who had
considered operators like eq. (1) long ago \cite{7} for raising this point and
for discussions.}  It thus follows rather conclusively that
the specific range of values of $\rm \delta m^2$ reported by
SuperKamiokande cannot reasonably
be accommodated within the standard model, even
with the inclusion of quantum gravity, and thus there \underline{must}
exist new physics beyond the standard model.

{\bf 3.  The Nature of New Physics:}  We now go further and turn to the
second point about the nature of the new physics, suggested by the SK
result.  The only reasonable way to understand a mass for the neutrino
or $\rm \delta
m^2$, as observed, it seems to me, is to introduce a right-handed (RH)
neutrino $(\rm \nu_R)$ and utilize the see-saw mechanism (described
below).\footnote[2]{The alternative of giving a Majorana mass to $\rm
\nu_L$ through renormalizable interaction
by introducing a $\rm SU(2)_L$ Higgs-triplet $\xi$ and
choosing the corresponding (Yukawa coupling) $\times$ (VEV of $\xi$) to
be nearly (1/10 - 1/30)eV seems to be rather arbitrary.}
This in turn has far-reaching implications.  The existence of a
RH neutrino becomes compelling by extending the SM symmetry to include
either SU(4)-color or the left-right symmetric gauge-structure
$\rm SU(2)_L \times SU(2)_R$, \cite{2}.
Thus the SK result motivates, on observational ground,
the route to higher unification
via the gauge-structure:
\begin{equation}
{\rm G(224) = SU(2)_L \times SU(2)_R \times SU(4)^C}.
\end{equation}
This is the minimal extension of the SM that specifies all
quantum numbers (given a representation), quantizes electric charge and
introduces $\nu_R$.
With respect to $\rm G(224)$, quarks and leptons of a given family fall
into the neat pattern \cite{2}:
\begin{equation}
{\rm
F^e_{L,R} = \left[
\begin{array}{llll}
u_r & u_y & u_b & \nu_e \\
d_r & d_y & d_b & e^-
\end{array}
\right]_{L,R}}
\end{equation}
with the transformation properties ${\rm F^e_L = (2,1,4)}$, and ${\rm F^e_R =
(1,2,4);}$ likewise for the $\mu$ and the $\tau$-families.  We see that
the RH neutrino $(\rm \nu_R)$ arises as the fourth color partner of the
RH up-quarks and, also, as the left-right conjugate partner
of the LH neutrino $\rm(\nu_L)$.
It is worth noting that the symmetry G(224), subject to
L-R discrete symmetry [2,8], possesses
some additional advantages, even without being embedded into a simple
group like $\rm SO(10)$ \cite{5} or $\rm E_6$ \cite{9}.  These
include:  (i) inclusion of all members of a family into one multiplet,
(ii) quark-lepton unification through SU(4)-color, (iii)
quantization of electric charge, mentioned above, (iv) spontaneous
violations of parity \cite{2,8} and of CP \cite{10}, (v) (B-L),
as a local symmetry whose spontaneous violation may be needed to implement
baryogenesis \cite{11}, (vi) a promising solution to
the strong CP problem in the context of supersymmetry \cite{12}, and
(vii) a possible resolution of the $\mu$-problem in the same context 
\cite{13}.
Embedding G(224) into SO(10), for which $\rm (F^e_L + \overline{F^e_R})$
yield the 16 of SO(10), would of course retain most of these
advantages, except possibly (vii).  Last, but not least, the symmetry
G(224) can emerge from strings with three chiral families (see e.g.
Refs. 14 and 15).  In this case,
the gauge coupling unification \cite{16} at string scale
would still hold \cite{17} even without having the covering SO(10),
below the string scale.\footnote[3]{Possible resolutions of a mismatch between
MSSM and string-unification scales by about a factor of 20 have been
proposed, including one that suggests two vector-like families $\rm (16 +
\overline{16})$ at the TeV-scale, that leads to semi-perturbative unification
and raises $\rm M_X$ to a few $\rm \times 10^{17}$ GeV\cite{18}; and also one
that makes use of string duality\cite{19} and allows for a re-evaluation of
$\rm M_{string}$ compared to that of Ref. [17].  In general, both ideas may
play a role.}  It is worth noting that in the
string context there is a distinct advantage if the preferred string
solution would contain G(224) rather than SO(10), because it appears
rather difficult to implement doublet-triplet splitting for
string-derived SO(10) so as to avoid rapid proton decay.\cite{20}  For
string-derived G(224) \cite{14}, on the other hand, the dangerous color
triplets are either projected out or naturally become superheavy.

{\bf 4.  The Mass of $\rm \nu_L^\tau$:}  I now turn to an estimate of
the masses of the light neutrinos, that are observed in the laboratory,
especially the $\rm \nu^\tau_L$, allowing for the existence of the RH
neutrinos $\rm (\nu_R's)$.  For this purpose, I will work with either
G(224) or its natural extension SO(10).  With a string or a GUT-origin, one can
motivate the symmetry-breaking scale for either G(224) or SO(10),
to be around $\rm M_{string}/10$, which
is nearly the (empirical) GUT-scale $\rm \approx 2 \times 10^{16} GeV$.

The amusing thing is that, in contrast to the case of the SM (eq.(1)),
now the mass of $\rm \nu^\tau_L$ comes out to be just in the right
range, so as to be relevant to the SK result.

The simplest reason for the known neutrinos to be so light $(\rm < 30
eV$ (say)) is provided by the so-called see-saw
mechanism \cite{3}.  It utilizes the fact that neutrinos being
electrically neutral can have two sources of mass:  (i) first, with both
$\rm \nu^i_L$ and $\rm \nu^i_R$, neutrino of the ith family would
naturally acquire a Dirac mass $\rm m(\nu^i_D)$ which would be related
to the up-flavor quark-mass $\rm (m_u, m_c$ or $m_t)$, depending upon the
Higgs representation (see below), by SU(4)-color.  (ii)  Second, since
RH neutrinos are standard model singlets they can acquire superheavy
Majorana masses $\rm (M_{iR})$, preserving the SM symmetry; by utilizing the
VEV of a suitable Higgs multiplet (call it $\Sigma)$, which would be
involved in breaking SO(10) or G(224) to the SM symmetry
G(213).  Before discussing the choice of $\Sigma$ and its coupling, let
us recall that a mass-matrix involving Dirac and superheavy Majorana masses, as
mentioned above, would diagonalize to yield three superheavy RH
neutrinos with masses $\rm M_{iR}$ and three light LH neutrinos with masses
\cite{3}:
\begin{equation}
\rm m(\nu^i_L) \approx m(\nu^i)^2_D/M_{iR}
\end{equation}
In writing this, we have neglected (for simplicity) possible off
diagonal mixings between different flavors.  Since we will be interested
in this note primarily in the mass of the heaviest one among the light
neutrinos (i.e. $\rm \nu^\tau_L)$, such mixings will not be so important.
(For a more general analysis, see e.g. Ref. 4 and 21).  Since the Dirac
masses enter quadratically into (5), and are highly hierarchical (e.g. $\rm
m_u:m_c:m_t \approx 1:300:10^5$), we expect, even allowing for a rather
large hierarchy (by successive factors of order 100, say) in $\rm
M_{iR}$, that the masses of the left-handed neutrinos will be light but
hierarchical $\rm (m(\nu^e_L) << m(\nu^\mu_L) << m(\nu^\tau_L))$.

The Higgs multiplet $\rm \Sigma$, mentioned above, and its conjugate
$\bar {\Sigma}$ (needed for supersymmetry), can either be in a symmetric
tensorial representation\cite{3} - i.e. $\rm (126_H$, $\rm
\overline{126}_H)$ of SO(10) or equivalently [(1,3,10), $\rm
(1,3,\overline{10})]$ of G(224) - or in the spinorial representation -
i.e. $\rm (16_H$, $\rm \overline{16}_{H})$ \cite{22} of SO(10) -
or equivalently in $\rm [(1,2,\overline{4})_H$, $\rm (1,2,4_{H}$)]
\cite{2} of G(224), like the quarks and the leptons.  For a
string-derived G(224), the L-R conjugate multiplets (like $\rm
(3,1,10)_H$ or $(2,1,\overline{4})_H$ etc.) should also exist if L-R discrete
symmetry (i.e. parity) is preserved in the Higgs-sector, following
string-compactification.  (In general, even if G(224) emerges as a gauge
symmetry, after compactification, and the spectrum of 3 chiral families
respect L-R discrete symmetry, the full spectrum need not.  See e.g.
Ref. 14, where the multiplets $\rm (1,2,4)_H$ and
$\rm (1,2,\overline{4})_{H}$ do
emerge, but not their (L-R) conjugates $\rm (2,1,4)_H$ and $\rm
(2,1,\overline{4})_{H}$.)

We first remark that, in string theory, the tensorial representations
$\rm 126_H$ and $\rm \overline{126}_{H}$, and likewise $\rm (1,3,10)_H$ and
$\rm (1,3,\overline{10})_H$, which can have renormalizable Yukawa
interactions with quarks and leptons,
are hard, perhaps impossible, to realize \cite{23}, and have not been
realized in any solution yet.  By contrast, the spinorial $\rm 16_H$ and
$\rm \overline{16}_{H}$, as also $\rm (1,2,4)_H$ and
$(1,2,\overline{4})_{H}$, do emerge quite simply in string-solutions
(see e.g. Ref. 14 for G(224) and Ref. 20 for
SO(10)).  Taking this as a good guide, and believing in the
string-origin of the effective theory just above the GUT-scale, we will
work only with the spinorial $\rm 16_H$ and $\rm \overline{16}_{H}$, or
equivalently with $\rm (1,2,4)_H$ and $(1,2,\overline{4})_{H}$.

The effective non-renormalizable interaction, involving these
multiplets, which we expect might be
induced by Planck-scale physics, and would give Majorana masses to the RH
neutrinos, are then\footnote[4]{We are not exhibiting the interactions of
$\rm (2,1,\overline{4})_{H}$ because, either it is absent (as in Ref. 14) or
has zero VEV.}
\begin{equation}
\rm {\cal L}_M (SO(10)) = \lambda^{ij}_R 16_i \cdot 16_j \overline{16}_H
\cdot \overline{16}_H/M_{P \ell} + hc
\end{equation}
\begin{equation}
\rm {\cal L}_M (G(224)) = \lambda^{ij}_R (1,2,4)_i (1,2,4)_j
(1,2,\overline{4})_H (1,2,\overline{4}_H)/M_{p \ell} + hc
\end{equation}
Here, i, j = 1, 2, 3, correspond respectively to e, $\mu$ and
$\tau$-families.  Note that in each case,
we have set the scale of the interaction to be
given by the reduced Planck mass, as in eq. (1).  Such effective
non-renormalizable interactions may well arise -- in part or dominantly
-- by \underline{renormalizable} interactions through tree-level exchange of
superheavy states, such as those in the string-tower (see remarks later).

Judging from the string-side, one naturally expects the VEVs of fields
which break GUT-like symmetries -- i.e. SO(10) or G(224) -- to the
standard model symmetry to be of order
$\rm M_{string}/(5 \; to \; 20) \approx 2-8 \times 10^{16} GeV$ [see,
e.g. Ref. 24 and 14], where $\rm M_{string} \approx 4 \times 10^{17}$
GeV.\cite{17}.  Interestingly enough, this is also nearly the GUT-scale
$\rm (M_{GUT} \approx 2 \times 10^{16} GeV)$, as
judged from the MSSM extrapolation of the three
gauge-couplings,\footnote[3]{} which should therefore represent the VEVs
of fields like $\rm < \overline{16}_H >$ or $\rm < (1,2,4)_H >$, which
break SO(10) or G(224) to the SM.  (For SO(10),
the VEV of $\rm < \bar{16}_H >$ may possibly
be somewhat larger than $\rm M_{GUT}$, because $\rm < \bar{16}_H >$
breaks SO(10) to SU(5) rather than the Standard model.)  Thus, both from
the viewpoint of connection with string theory, as well as comparison
with the MSSM unification-scale, we expect the
VEV's of the respective fields to be given by:
\begin{equation}
\rm For \; SO(10): \quad < 16_H > \; = \; < \overline{16}_H >
\; \approx 3 \times 10^{16} \; GeV.\eta
\end{equation}
\begin{equation}
\rm For \; G(224): \; < (1,2,\overline{4})_H > \; = \;
< (1,2,4)_H > \; \simeq \; 3 \times 10^{16} \; GeV.\eta
\end{equation}
with $\rm \eta \approx 1/2 \; to \; 2$, being the most plausible range.
Thus, using (6) -- (7) and (8) -- (9), for either SO(10) or G(224),
the Majorana masses of the RH neutrinos are given by:
\begin{equation}
\rm M_{iR} \approx \lambda_{ii} \frac{(3 \times 10^{16}GeV)^2}{2 \times
10^{18} GeV} \eta^2
\newline \approx \lambda_{ii} (4.5 \times 10^{14} GeV) \eta^2
\end{equation}
In writing (10), we have ignored the effects of off-diagonal mixing.
This is justified, especially for the third family, if we assume, as we
do, that the Majorana couplings are family-hierarchical, $\rm
\lambda_{33}$ being the leading one, somewhat analogous to those that
give the Dirac masses.

Now using SU(4)-color and the Higgs multiplet $\rm (2,2,1)_H$ for G(224) or
equivalently $\rm 10_H$ for SO(10), one obtains the relation
$\rm m_\tau (M_X) = m_b (M_X)$, which is known to be successful.
Thus, there is a good reason to believe that the third family gets its masses
primarily from the $\rm 10_H$ or equivalently $\rm (2,2,1)_H$,
which automatically gives the same
Dirac mass to the quark and the lepton of a given flavor.  (In the
context of SUSY, one would need two 10's or two (2,2,1)'s, or effective
non-renormalizable operators, to
induce CKM mixings).  In turn this implies:
\begin{equation}
\rm m(\nu^\tau_D) \approx m_{top}(M_X) \approx (100-120) GeV
\end{equation}
combining (10) and (11) via the see-saw relation (5), we obtain:
\begin{equation}
\rm m(\nu^\tau_L) \; \approx \; \frac{(100 GeV)^2 (1 \; to \;
1.44)}{\lambda_{33}(4.5 \times 10^{14} GeV) \eta^2}
\approx (1/45) eV (1 \; to \; 1.44)/\lambda_{33} \eta^2
\end{equation}
Now, considering that we expect $\rm m(\nu^\mu_L) << m(\nu^\tau_L)$ (by
using eq. (5)), and assuming that
SuperKamiokande observation represents $\rm \nu^\mu_L
\rightarrow \nu^\tau_L$-oscillation, so that the observed $\rm
\delta m^2 \approx 10^{-2} to 10^{-3} eV^2$ corresponds to $\rm
m(\nu^\tau_L)_{obs} \approx 1/10 \; to \; 1/30 eV$,
it seems {\it truly remarkable}
that the expected magnitude of $\rm m(\nu^\tau_L)$, given by eq. (12),
is just about what is observed, if $\rm \lambda_{33} \eta^2 \approx 1 \; to
\; 1/4$.  Such a range for $\rm \lambda_{33}\eta^2$ seems most plausible and
natural (see remarks below).  This observation regarding the agreement
between the expected and the observed value of $\rm \delta m^2$ (in this
case $\rm m(\nu^\tau_L))$, in the context of the ideas mentioned above,
is the main point of this note.

We remark that this agreement has come about without making any
parameter unnaturally small or large.  In particular, the effective Majorana
coupling of the third family $\rm (\lambda_{33})$ is needed to be
nearly one or order one for this agreement to hold.  One is tempted to
compare with the top-Yukawa coupling $\rm(h_{top})$ which is also nearly
one.  This common feature regarding maximality of the dimensionless
couplings associated with the third family (i.e. $\rm \lambda_{33} \sim
h_{top} \sim 1)$ may well find its explanation in the
context of string solutions for which such couplings
may be given just by the gauge
coupling [e.g. $\rm h_{top} = \sqrt{2} g \approx 1$,
[see e.g. Ref. \cite{24}] and are thus of order one\footnote[5]{Although
$\rm \lambda_{ij}$ are associated with effective non-renormalizable
couplings, as mentioned before, they may well arise, in part or
dominantly, through the exchange of superheavy states $\{ \phi_\alpha
\}$ (such as those in the string-tower or just below string-scale),
if these possess Yukawa
couplings of the form $\rm h_{i \phi} 16_i \overline{16}_H \phi$,
together with invariant mass-term $\rm (M_\phi \phi \phi + hc)$.  If
$\rm h_{(i \phi)}$ are family-hierarchical with $\rm h_{3 \phi}$ being
maximal (i.e. {\cal{O}}(1) like $h_{top}$) and leading, $\rm
\lambda_{ij}'s$ would also be hierarchical, with $\rm \lambda_{33} (\rm = 
h^2_{3 \phi} (M_{Pl}/M_{\phi}))$ being maximal ($\cal O$(1)) and leading.},
while those associated with the second and the first
families are progressively smaller, because, subject
to string symmetries and selection rules, they are induced only at the
level of higher dimensional operators utilizing VEV's of fields which
are small (by nearly factor of 10) compared to the string-scale.  In
addition to $\rm \lambda_{33}$, the value of
$\rm m(\nu^\tau_L)$ depends on two other parameters - i.e.
on the Dirac mass $\rm m(\nu^\tau_D)$ (see eq. (5)) and on the VEV of
$\rm < \overline{16}_H >$ or $<(1,2,4)_H>$, and thus on
$\eta^2$ (See eqs. (8)/(9), (10) and (5).)  As regards the Dirac mass,
the use of SU(4)-color plays a crucial role in that it
enables one to determine $\rm
m(\nu^\tau_D)$ fairly reliably from $\rm m_{top}$, extrapolated to the
GUT-scale (see eq. (11)).  As regards determining the VEVs of fields mentioned
above, the use of string as well as GUT-related ideas yield nearly the
same value for the VEV of $\rm < 16_H >$ or $\rm <(1,2,\bar{4})_H>$, within a
factor of 2 to 4, which is reflected in the uncertainty in $\rm \eta
(\approx 1/2 \; to \; 2)$ (see eqs. (8)/(9)).  It is for these reasons that
the value of $\rm m(\nu^\tau_L)$ obtained in eq. (12), with $\rm
\lambda_{33} \eta^2 \approx 1 \; to \; 1/4$, seems most plausible.
\footnote[6]{Note that $\rm m(\nu^\tau_L)$ depends in fact only on the
{\it product} $\rm \lambda_{33} \eta^2$ (see eqs. (10) and (12)). A more
precise understanding of $(\rm \lambda_{33} \eta^2)$ and thereby of $\rm
m(\nu^\tau_L)$ would of course
still need a sharpening of an understaning of $\eta$, as well as of $\rm
\lambda_{33}$, e.g. in the context of string-solutions [see remarks in
Footnote 5].}

Together with the result $\rm \delta m^2 \simeq 10^{-2} - 10^{-3} eV^2$,
the SuperKamiokande group reports another puzzling feature that $\rm
\nu_\mu \rightarrow \nu_\tau$ (or $\rm \nu_X$) -oscillation angle is
nearly maximal - i.e. $\rm \sin^2 2 \theta > 0.8$.  Ordinarily, such
large oscillation angle is attributed to nearly degenerate masses of the
$(\rm \nu_\mu - \nu_\tau)$ or $\rm (\nu_\mu - \nu_X)$ systems, as many
authors in fact have.  In this case, the large oscillation angle
is attributed almost entirely to a large or maximal
mixing in the mass eigenstates of the respective neutrinos.
However, considering that nearly degenerate
masses for the light neutrinos seem to be rather unnatural in the context of
the see-saw formula, Babu, Wilczek and I have very recently
observed \cite{4} that such degeneracy is not even needed to
obtain large oscillation angle.
By {\it combining} the contributions from the mixing angle
of the neutrinos (i.e. $\rm \nu^\mu_L - \nu^\tau_L$) with that from the
charged leptons ($\mu - \tau$), and by following familiar ideas on the
understanding of Cabibbo-like quark-mixing angles,
one can in fact obtain, quite simply
and naturally, large $(\nu_L^\mu - \nu_L^\tau)$ oscillation angle, as observed,
{\it in spite of a highly
non-degenerate} $\nu^\mu - \nu^\tau$ {\it system}
\footnote[7]
{Preliminary aspects of this joint work \cite{4} were presented by
J.C. Pati at the $\rm \nu-98$ Conference held at Takayama, June 6,
1998.}  
e.g. with $\rm
m(\nu^\mu_L)/m(\nu^\tau_L) \approx 1/10 - 1/20$.  
Briefly, a simple and plausible origin of the large mixing angle
is as follows. If one assumes that the lighter eigenvalue for a
hierarchical $\rm 2 \times 2$-system arises entirely or primarily by the
off-diagonal mixing of the (would-be) light with the heavier state (as
in a symmetrical see-saw type mass matrix),
one obtains the familiar square root-formula\cite{wwf} for the mixing
angle, like $\rm \theta_{d,u} \approx (\sqrt{m_d/m_s},
\sqrt{m_u/m_c})$, and the Cabibbo angle is obtained by combining
$\theta_d$ with $\theta_u$, allowing for a relative phase between them.
Regardless of the phase, such an expression for the Cabibbo angle is 
known to be fairly successful (to better than 30 \%). Assuming analogous 
mass-matrices for the $\nu_\mu -
\nu_\tau$ system (Dirac and Majorana) as well as for the charged leptons
($\mu-\tau$), one obtains, ignoring CP violation (and assuming the exact
see-saw form for each of the three matrices):
$\rm \theta_{osc}(\nu_\mu-\nu_\tau)=\theta(\nu^\mu_L-\nu^\tau_L)\pm
\theta(\mu-\tau)\approx [m(\nu^\mu_L)/m(\nu^\tau_L)]^{1/2} \pm 
[m_\mu/m_\tau]^{1/2} \approx 0.31 \pm 0.25 \approx 0.56 \, or \, 0.06$, where
we have put $m(\nu^\mu_L)/m(\nu^\tau_L) \approx 1/10$. This yields,
choosing a positive relative sign between the two mixing angles, $\sin^2 2
\theta_{osc}\approx 0.8$. {\it In
short, a large oscillation angle can arise quite plausibly, without near 
degeneracy and without large mixing in the mass eigenstates of the 
neutral and the charged leptons.} Various sources of departures from the
simple square root formula for the mixing angle (corresponding to
departures from exactly symmetrical see-saw mass matrices),
which can lead to even larger oscillation angles (for ${m(\nu^\mu_L) / 
m(\nu^\tau_L)} \approx 1/10-1/20)$, are discussed in Ref\cite{4}:
In this case, $\rm \nu_e - \nu_\mu$ - oscillation can become relevant
to the small angle MSW explanation\cite{25}
of the solar neutrino-puzzle.  I refer the reader to Ref. 4 for a full
discussion of this explanation of the large oscillation angle for the
$\nu_\mu - \nu_\tau$ system, with hierarchical masses for the neutrinos.
The purpose of the present note has
primarily been to emphasize the implications
of the observed {\it magnitude} of
$\rm \delta m^2$- or equivalently, in our case of $\rm m(\nu^\tau_L)$,
on the nature of new physics.

{\bf 5.  Link Between Neutrino Masses and Proton Decay}.

Proton decay is one of the hallmarks of grand unification
[\cite{2},\cite{6}].  As discussed here, light neutrino masses $\rm (<<
m_{e, \mu, \tau})$ are also an important characteristic of symmetries 
such as
G(224) and SO(10), assuming that they are supplemented by the see-saw
mechanism.  Ordinarily, except for the scale of new physics, involved in
the two cases, however, proton decay, especially its decay modes, are
considered to be essentially unrelated to the pattern of neutrino
masses.  In a recent paper, Babu, Wilczek and I noted that, contrary to
this common impression, in a class of supersymmetric unified
theories such as SUSY SO(10) or SUSY G(224), there is likely to be an {\it 
intimate link} between the neutrino masses and proton decay\cite{21}
\footnote[8]{ The link is most compelling for the case of 
$\overline{126_H}$ giving 
Majorana masses to the RH neutrinos.  It becomes compelling also for the
case of ($\overline{16_H},16_H$), serving the same purpose, when one
attempts to understand not only the masses but also the CKM mixings of
quarks \cite{21}.}.  This is
because, in the process of generating light neutrino masses via the
see-saw mechanism, one inevitably introduces a new set of color-triplets
(unrelated to electroweak doublets), with effective couplings to quarks
and leptons, which are related to the superheavy Majorana masses of
the RH neutrinos (see eqs. (6) and (7)).  Exchange of these new
color-triplets give rise to a new set of d=5 proton decay operators,
which are thus directly related to the neutrino-masses.  Assuming that
$\rm \nu_e - \nu_\mu$ oscillation is relevant to the MSW explanation of
the solar neutrino puzzle, so that $\rm m(\nu^\mu_L) \approx 3 \times
10^{-3} eV$, which corresponds to $\rm M(\nu^\mu_R) \approx 2 \times 10^{12}$
GeV, the strength of the new d=5 operators turns out to be just about
right $(\tau_P \approx 10^{32.5\pm2})$ yrs, for proton decay to be
observable at SuperKamiokande\cite{21}.

The flavor-structure of the new d=5 operators are, however, expected to
be distinct from those of the standard d=5 operators, which are related
to the highly hierarchical Dirac masses of quarks and leptons.  In
contrast to the standard d=5 operators, the new ones can lead to
prominent (or even dominant) charged lepton decay modes, such as
$\ell^+ \pi^o, \ell^+ K^o$ and $\ell^+ \eta$, where $\ell$ = e or $\mu$,
even for low or moderate values of $tan \beta \leq$ 10.
The intriguing feature thus is that owing to the underlying SO(10)
or just SU(4)-color symmetry, {\it proton decay operator knows about
neutrino masses and vice versa}.

With a {\it maximal} effective Majorana-coupling for the third family (i.e.
$\rm \lambda_{33} \sim {\cal{O}}(1))$, as suggested here, that corresponds
to $\rm M_{3R} \approx (few \times 10^{14} GeV)$ (see eq. (10)) and
thereby to $\rm m(\nu^\tau_L)$ agreeing with the SuperKamiokande value
(eq. (12)), one might however worry that proton may decay too fast, because of
an enhancement in the new d=5 operators, relative to that considered in
Ref. 21.  It turns out, however, that
because $\rm \tau^+$ is heavier than the proton and because $\rm
\bar{\nu}_\tau K^+$ mode receives a strong suppression-factor from the
small mixing angle associated with
the third family $(V_{ub} \approx 0.002 - 0.005)$, a
maximal Majorana-coupling of the third family $\rm (\lambda_{33} \sim
{\cal{O}}(1))$, and thus $\rm m(\nu^\tau_L) \approx (1/10 - 1/30) eV$, is
perfectly compatible with present limit on proton lifetime\cite{4}.
With a family-hierarchical Majorana coupling - i.e. $\rm \lambda_{33}
\sim {\cal{O}}(10) \lambda_{23} \approx {\cal{O}}(10^2) \lambda_{22}$
etc. - $\rm \nu^\tau_L$ and $\rm \nu^\mu_L$ - masses can be
relevant respectively to the atmospheric and the solar-neutrino-problems, yet
the new neutrino-mass related d=5 operator does not conflict with
proton lifetime.  They would still give observable rates for
proton-decay, with prominent charged lepton decay modes, involving at
least the second family (i.e. $\rm (\mu^+ \pi^o, \mu^+ K^o$ etc.), together
with $\rm \bar{\nu} K^+$ modes \cite{4}.
Observation of proton decay, together with prominence of charged lepton
modes, would thus be a double confirmation of both susy-unification
through G(224)/SO(10), as well as of the ideas on neutrino masses in
this context.

{\bf 6.  Concluding Remarks and a Summary:}
As noted in the introduction and the
subsequent sections, the impressive result of SuperKamiokande clearly
has far-reaching implications on the nature of new physics.  These are
summarized below and some remarks are added:

{\bf (1) The Right-Handed Neutrino:  A New Form of Matter}:
As noted in the introduction, the most reasonable
explanation for the neutrino mass-scale observed at SuperKamiokande needs
a RH neutrino $(\rm \nu_R$).  Many in the past, motivated by the
possible masslessness of neutrinos, have preferred to view the neutrino
as an "odd ball," believing that it is the messenger that nature is
intrinsically left-right asymmetric (parity-violating).  This is
reflected by the two-component neutrino hypothesis of Lee, Yang, Landau
and Salam, as well as by the hypothesis of the grand
unification-symmetry SU(5).  The SuperKamiokande result
(especially its value for $\rm \delta m^2)$ clearly suggests, however,
that that is in fact not the case.  {\it Neutrino is "elusive" but not
an odd ball after all.}  It has its RH counterpart (one for each flavor)
just like all the other fermions.

Nevertheless, the neutrino has a {\it unique character}.  It is the only
fundamental fermion, among the members of a quark-lepton family, that is
electrically neutral (not counting possible SUSY gauge matter such as
photino or gluino).  Therefore, it is the \underline{only fermion} that
can acquire both a Dirac mass $\rm (\Delta F = \Delta L = 0)$, combining $\rm
\nu_R$ and $\rm \nu_L$, and a Majorana mass for either $\rm \nu_R$ or
$\rm \nu_L$ $\rm (\Delta F = \Delta L =
\Delta (B-L) = \rm 2)$, conserving electric charge.  The Majorana masses
of the RH neutrinos can be superheavy, because they do not break the
Standard model symmetry.  As mentioned before,
this unique character of possessing both a
Dirac and a superheavy Majorana mass for the RH $\rm \nu_R$ allows the
LH neutrinos to be naturally light via the
see-saw mechanism.  The lightness of $\rm \nu_L$ is in fact a
reflection of the heaviness of $\rm \nu_R$.
By the same token, the light neutrinos know about
both mass-scales -- the Dirac and the Superheavy Majorana -- and thereby
simultaneously of the physics at the electroweak and the
string/GUT-scales.  In short, neutrino masses carry a {\it gold mine of
information} about the nature of new physics.

{\bf (2) Minimal Extension Needed of the Standard Model}:  In suggesting
the need for the RH neutrino, the SuperKamiokande result in turn
suggests, following discussions presented here, that the standard model
symmetry must be extended minimally to the symmetry-structure G(224) =
$\rm SU(2)_L \times SU(2)_R \times SU(4)^c$.
The need for $\rm SU(4)^c$ has been noted above and
is summarized below.  Strictly speaking, for an understanding of
$\rm (\delta m)^2$, as presented here, the extension of the SM symmetry to
just G(214) = $\rm SU(2)_L \times I_{3R} \times SU(4)^c$ would
suffice.\footnote[9]{For a string-origin of G(214), see Ref. \cite{26}.}
The further extension of G(214) to G(224) (that also quantizes electric charge
by replacing $\rm I_{3R}$ by $\rm SU(2)_R$)
may however be needed by some of the other considerations, listed in
Sec. 3, as well as those of fermion masses and mixings.  

{\bf (3) The Three Necessary Ingredients}:  Understanding the neutrino
mass-scale observed at SuperKamiokande, as discussed here, utilizes three
concepts in an essential manner.
They are:  (a) SU(4)-color that not only enforces $\rm
\nu_R$, but more importantly gives the Dirac mass of $\nu^\tau$, fairly
reliably, by relating it to the mass of the top quark (eq.
(11));\footnote[10]{It is, of course, possible that a string-derived
solution containing, for example, only G(2213) = $\rm SU(2)_L \times
SU(2)_R \times (B-L) \times SU(3)^C$ or G(2113) = $\rm SU(2)_L \times I_{3R}
\times (B-L) \times SU(3)^C$ \cite{24}, or flipped SU(5) $\times$
U(1)\cite{27}, all of which yield RH neutrinos,
may still relate $\rm m(\nu^\tau_D)$ to $\rm m_{top}$ at string-scale. 
This comes about because such a solution still remembers its origin
through SU(4)-color or SO(10).  Here, I am only discussing the minimal {\it
underlying symmetry} needed to remove arbitrariness in the choice of
$\rm m(\nu^\tau_D)$, which appears to be
SU(4)-color.} (b) String/GUT-scale physics
that determines the Majorana mass of the RH
tau-neutrino (subject to maximality of the effective coupling) (eqs.
(8)-(10)); and (c) the see-saw relation (eq. (5)).
Given the sensitivity of the final result to both the Dirac mass and the
VEV that determines the Majorana mass,\footnote[11]{Because of the
quadratic dependence of $\rm m(\nu^\tau_L)$ on both the Dirac mass $\rm
m(\nu^\tau_D)$ and the VEV of $\rm (1,2,4)_H$ or $\overline{16}_H$, that
determines the Majorana mass, with error in either one by a factor of 10
(say), one could have been off by orders of magnitude in the final
answer.} the agreement of the expected value with the observed one (for
most plausible values of $\rm \lambda_{33}\eta^2 \approx {\cal{O}}(1)$)
seems to suggest the correctness of each of the three ideas.

{\bf (4)  Selecting the Route to Higher Unification}:  Unlike proton
decay, which can probe directly into the full grand unification symmetry
(including gauge transformations of $\rm q \rightarrow \overline{q}$ and
$\rm q \rightarrow \overline{\ell}$), neutrino physics probes
directly into  $SU(2)_L \times SU(2)_R \times SU(4)^c$,
but not necessarily beyond.  For example, the results discussed here, such as
determinations of the Dirac and the Majorana mass of the $\tau$ neutrino
utilize only G(224), but not the full SO(10).\footnote[12]{The scale of
the VEV determining the Majorana mass assumes the relevance of
string/GUT scale physics.  But that can hold in a string theory, even if
it gives, after compactificatin, only G(224) and not the full SO(10)
(See also remarks in footnote 3).}  They have also utilized
supersymmetry, at least indirectly, because without it, there would be
no rationale for the use of string-GUT-related scale for the VEV of
$\rm \overline{16}_H$ or $\rm (1,2,\overline{4})_H$.

At the same time, by providing clear support for G(224), the SK result selects
out SO(10) or $E_6$ as the underlying grand unification symmetry, rather
than SU(5).  Either SO(10) or $\rm E_6$ or
both of these symmetries ought to be relevant at
some scale, and in the string context, that may, of course, well be in
higher dimensions, above the compactification-scale, below which there
need be no more than just the G(224)-symmetry.  If, on the other hand,
SU(5) were regarded as a fundamental symmetry, first, there would be no
compelling reason, based on symmetry alone, to introduce a $\rm \nu_R$,
because it is a singlet of SU(5).  Second, even if one did introduce
$\rm \nu^i_R$ by hand, the Dirac masses, arising from the coupling $\rm
h^i \overline{5}_i < 5_H > \nu^i_R$, would be unrelated to the up-flavor
masses and thus rather arbitrary (contrast with eq. (11)).  So also
would be the Majorana masses of the $\nu^i_R$'s, which are
SU(5)-invariant and thus can even be of order Planck scale (contrast
with Eq. (10)).  This would give $\rm m(\nu^\tau_L)$ in gross conflict
with the observed value.  We thus see that the SK result clearly disfavors
SU(5) as a fundamental symmetry, with or without supersymmetry.

It is worth noting that the precision LEP-data, exhibiting
coupling unification\cite{28},
as also proton-decay searches \cite{29}, are known
to disfavor non-supersymmetric grand unification, but are compatible
with either SUSY SU(5) or SUSY SO(10).  It is thus interesting that
the neutrino data \cite{1} revises this conclusion in a major way, by
disfavoring SUSY SU(5), and selecting out either string-derived SUSY
G(224), or SUSY SO(10).

In summary, it seems that the single discovery of atmospheric
neutrino-oscillation has brought to light the existence of the
right-handed neutrino and has reinforced the ideas of SU(4)-color,
left-right symmetry and see-saw.  The agreement between the
simplest estimate of the mass of the tau-neutrino, presented here, and
the``observed value'' suggests the correctness of these three
ideas. Simultaneously, it suggests the relevance of the
string/GUT-scale-symmetry-breaking, as opposed to
intermediate or TeV-scale breaking of (B-L).  {\it Any symmetry}
containing G(224) = $\rm SU(2)_L x \times SU(2)_R \times SU(4)^c$, such
as SO(10) or $\rm E_6$, would of course possess the same desirable
features as regards neutrino physics, as G(224).  Given the wealth of
insight already provided by the SuperKamiokande result, one looks
forward eagerly to further revelations of deeper physics in the coming
years from the neutrino-system through the many existing and the
forthcoming facilities, involving atmospheric, solar and
accelerator neutrinos.  In particular, one would like a clarification of
whether the SK result is observing $\rm \nu^\mu_L - \nu^\tau_L$ (as
assumed here) as opposed to $\rm \nu^\mu_L - \nu_X$ oscillation, and
whether the resolution to the solar neutrino-problem would favor the MSW
solution (supported here) as opposed to $\rm \nu_e - \nu_X$ or vacuum
oscillation.  One of course also looks forward to learning much about
further aspects of unification from searches for proton decay,
which, as we saw \cite{21} \cite{4}, is intimately related to neutrino 
masses, because of SU(4)-color and supersymmetry.

\vspace{1.5mm}

{\bf Acknowledgement:}  The research is supported in part by NSF Grant
No. Phy-9119745, and in part by the Distinguished Research Fellowship
awarded by the University of Maryland.  I wish to thank the Organizers
of the Neutrino-98 Conference, especially Y. Totuska, Y. Suzuki and T.
Kajita, for inviting me to speak at the Conference and for their kind
hospitality.  I have greatly benefitted from discussions and
communications with Schmuel Nussinov, Qaisar Shafi, Steven Weinberg, and
Edward Witten, and especially from collaborative discussions with Kaladi
S. Babu and Frank Wilczek.

\end{document}